\documentclass[prl,aps,twocolumn,amsmath,amssymb,showpacs]{revtex4}
\usepackage{graphicx}
\usepackage{bm}
\begin{document}

\title{Fluctuation relation and pairing rule for Lyapunov exponents of inertial particles in turbulence}

\author{Itzhak Fouxon$^1$}
\author{P\'eter Horvai$^2$}

\affiliation{$^1$ Racah Institute of Physics,
  Hebrew University of Jerusalem,
  Jerusalem 91904,
  Israel}
\affiliation{$^2$ Mathematics Institute, University of Warwick,
Coventry CV4 7AL, UK}

\begin{abstract}

We study the motion of small particles in a random turbulent flow assuming linear law of friction.
We derive a symmetry relation obeyed by the large deviations of the finite time Lyapunov exponents in the phase
space. The relation applies when either the statistics of the strain matrix is invariant under the transposition
or when it is time-reversible. We show that, as a result, the Lyapunov exponents come in pairs which sum is equal
to minus the inverse relaxation time of the particles. We use the pairing to consider the Kaplan-Yorke
dimension of the particles attractor in the phase space. In particular, the results apply to case of the flow
which is white noise in time.

\end{abstract}


\maketitle

Recently much interest was raised by the Gallavotti-Cohen type fluctuation relations \cite{GallavottiCohen} for the dissipative dynamics, see e. g. \cite{Gallavotti} and references therein. The relations describe the large deviations of different observables and apply to the chaotic dynamics with time-reversal symmetry \cite{Gallavotti}. An example is the relation on the entropy function describing the large deviations of the finite time Lyapunov exponents \cite{BFF,review}. It is natural to inquire whether other symmetries of the dynamics also generate the corresponding fluctuation relations. Here we answer this question affirmatively in the
context of the chaotic dynamics describing the motion of inertial particles in a turbulent flow. We show
that two different {\it statistical} symmetries of the strain matrix of the flow lead to a new fluctuation relation for the finite time Lyapunov exponents. This relation implies that the spectrum of the Lyapunov exponents has a reflection point: the exponents come in pairs which sum is equal to minus the inverse relaxation time of the particles.

The behavior of small inertial particles in turbulence is now enjoying much attention from the researchers, see
e. g. \cite{Shaw,Nature,Piterbarg,engineering,MWB,DFTT,Horvai,Jeremie,MaxRil,HorvaiFouxon}.
The problem has many applications including rain formation in clouds \cite{Shaw,Nature}, ocean physics \cite{Piterbarg} and engineering \cite{engineering}. Theoretical study of the problem mostly involved modeling turbulence
by a white noise in time, Gaussian velocity field, known as Kraichnan model \cite{Piterbarg,review}. Even in
that case theoretical study
is rather difficult and analytic results were mainly obtained for the one-dimensional case \cite{MWB,DFTT}. In
particular, the spectrum of the Lyapunov exponents is unknown in dimension larger than one and probably cannot be found analytically \cite{Horvai}. In such a situation numerical studies are necessary \cite{Jeremie,Horvai}. The pairing of the Lyapunov exponents proved here for a class of flows that includes Kraichnan model is expected to be a useful tool in such studies.

Let us briefly review the study of the pairing of the Lyapunov exponents for dynamical systems.
The pairing is a well-known property of the Hamiltonian systems: the exponents come in pairs
with zero sum, see e. g. \cite{EckmannRuelle}. This is a consequence of the symplectic property of the linearization of the flow, see \cite{EckmannRuelle} and below. The first observation of the pairing for the dissipative dynamics is due to U. Dressler \cite{Dressler}. It was shown in \cite{Dressler} that the addition of the linear friction term to the Hamiltonian dynamics merely shifts the (zero) reflection point of the spectrum by minus half the friction coefficient. Later, the relation appeared in the
context of the thermostatted systems which are Hamiltonian systems with an added non-linear friction term \cite{Dorfman,Ruelle}. When the friction term is such as to make the kinetic energy of the system constant (the so-called isokinetic thermostat \cite{Ruelle}), it was shown by C. P. Dettmann and G. P. Morriss in \cite{DettmannMorriss} that,  after the exclusion of two zero Lyapunov exponents, the remaining Lyapunov exponents
obey the conjugate pairing rule. The exponents come in pairs which sum to the finite-time average of the phase space contraction. Both Dressler and Dettmann-Morriss relations apply for the finite time Lyapunov exponents and express
the fact that Jacobi matrix of the flow is {\it conformally symplectic} -- it transforms simply under the similarity transformation by the symplectic tensor, see \cite{DettmannMorriss,Ruelle} and below. On the other hand, it was argued that for the isoenergetic thermostat that conserves the total rather than the kinetic energy, the exact pairing of the finite time Lyapunov exponents should not hold \cite{Bonnetto}. Still, the pairing may be true in the limit of large times (it is almost certainly true for both large times and large number of particles when equivalence
between the isokinetic and isoenergetic thermostats is expected) \cite{Bonnetto}. The isoenergetic thermostat is an example of the situation where Jacobi matrix is not conformally symplectic and yet the pairing is expected to hold asymptotically. To treat such cases different techniques are needed.  For the case of a thermostatted system where the interactions can be described by white noise, the use of Fokker-Planck equation allowed to prove the pairing rule in \cite{TaniguchiMorriss}. Here we suggest a different approach to pairing in different situations that does not involve the assumption of the white noise. We note that to prove the conjugate pairing rule for the infinite time Lyapunov exponents it is sufficient to show that Jacobi matrix is only {\it statistically} conformally symplectic. Here statistics is defined by sampling the initial points with the natural measure. Indeed, after it is proved that Jacobi matrix is
statistically conformally symplectic, one can use the multiplicative ergodic theorem by Oseledec \cite{Oseledec}
that ensures that the part of the matrix defining the Lyapunov exponents does not fluctuate in the limit of the large times. Thus in the limit of the large times the statistical property brings a deterministic pairing rule for the infinite time Lyapunov exponents.

We pass to the analysis. We consider the motion of a small spherical particle immersed in a fluid. We assume the $d-$dimensional flow
of the fluid $\bm u(\bm r, t)$ to be smooth below some scale $\eta$ (if $\bm u$ is a developed turbulent
flow then $\eta$ is the Kolmogorov scale). Due to the finite inertia the particle deviates from the carrying flow.
Provided the particle is sufficiently small and heavy, the drag by the flow is well described by the linear, Stokes' law of friction, see below and \cite{MaxRil} for the validity conditions. Designating
the particle position and velocity by ${\bm r}(t)$ and ${\bm v}(t)$, Newton's law reads
\begin{equation}
\label{Newton}
  \dot{\bm r}
=
  \bm v
\,\,,\qquad
  \dot{\bm v}
=
  -\frac{\bm v - \bm u(\bm r(t),t)}{\tau}
\,.
\end{equation}
Here $\tau = (2/9)(\rho_0/\rho)(a^2/\nu)$, where $\rho_0$ and $a$ are the
particle density and radius, while $\rho$ and $\nu$ are the fluid's density and
kinematic viscosity. Above the particle radius should be much smaller than $\eta$,
the Reynolds number associated with the particles' motion relative to the flow should
be small and the particle should be sufficiently heavy for effects like added mass
to be negligible, see \cite{MaxRil} for the details. The trajectories $\bm X(t, \bm x)$ in
the phase space of Eq.~(\ref{Newton}) obey
\begin{eqnarray}&&
\frac{\partial \bm X(t, \bm x)}{\partial t}=\bm V\left(\bm X(t, \bm x) , t\right),\ \
\bm X(0, \bm x)=\bm x, \label{dynamics}
\end{eqnarray}
where $\bm V\equiv (\bm v, -[\bm v-\bm u(\bm r, t)]/\tau)$. Note that the flow in the phase space is
dissipative: $\nabla\cdot\bm V=-d/\tau$. Below we shall assume that $\bm u(\bm r, t)$ is
a random flow with stationary statistics. As a concrete example we shall use
Kraichnan model where turbulence is modeled by a Gaussian velocity field $\bm u(\bm r, t)$
with zero average and pair correlation
\begin{eqnarray}&&
\!\!\!\langle u_i(\bm r_1, t_1)u_j(\bm r_1, t_2)\rangle=\delta(t_2\!-\!t_1)\left[V_0\delta_{ij}\!-\!K_{ij}(\bm r_2-\bm r_1)\right].
\nonumber
\end{eqnarray}
Above the tensor $K_{ij}(\bm r)$ is assumed to have finite second derivatives at $\bm r=0$ in conformance
with the assumption that the flow is smooth. Due to the randomness of $\bm u(\bm r, t)$ the flow in the
phase space is chaotic for typical realizations and it can be characterized by the spectrum of the Lyapunov exponents which we introduce now. We consider the Jacobi matrix $W_{ij}(t, \bm x)\equiv \partial_j X_i(t, \bm x)$ determining the evolution of infinitesimal volumes in the phase space.
Consider as an example the distance $(\bm R, \bm V)$ between two infinitesimally close trajectories in the phase space that satisfies
\begin{eqnarray}&&
\frac{d\bm R}{dt}=\bm V,\ \ \frac{d\bm V}{dt}=-\frac{\bm V-\sigma\bm
R}{\tau}. \label{separation}
\end{eqnarray}
where $\sigma_{ij}=\nabla_j u_i(\bm r(t), t)$ is the strain matrix evaluated at the trajectory of the particle.
The solution can be written as $[\bm R(t), \bm V(t)]=W(t)[\bm R(0), \bm V(0)]$. Jacobi matrix obeys the corresponding equation
\begin{eqnarray}&&
\frac{d W}{dt}=\begin{pmatrix}
    0 & 1 \\
    \frac{\sigma}{\tau} & -\frac{1}{\tau}\\
  \end{pmatrix} W, \ \ \ \  W(0)=I. \label{evolutionequation}
\end{eqnarray}
where each entry in the stability matrix above is a $d\times d$ matrix and $I_{ij}=\delta_{ij}$ is the $2d-$dimensional unit matrix.  According to Oseledec theorem \cite{Oseledec},
the evolution of distances has well-defined exponential behavior at large times corresponding to the existence
of the finite matrix limit $\lim_{t\to\infty}\ln W^T(t, \bm x)W(t, \bm x)/2t$. Assuming that the system is ergodic,
the eigenvalues $\lambda_i$ of the limiting matrix are the same for almost every realization of the velocity and they can be considered independent of $\bm x$. These eigenvalues define the Lyapunov spectrum of the system. Conventionally we arrange the exponents in the non-decreasing order, $\lambda_i\geq \lambda_{i+1}$.
We introduce the finite time Lyapunov exponents $\mu_i(t, \bm x)$ as the eigenvalues of $[\ln W^T(t, \bm x)W(t, \bm x)]/2t$, so that $\lim_{t\to \infty}\mu_i(t, \bm x)=\lambda_i$ (again $\mu_i\geq \mu_{i+1}$). The following constraint follows from $\nabla\cdot \bm V=-d/\tau$:
\begin{eqnarray}&&
\sum\mu_i(t)=\sum\lambda_i=-\frac{d}{\tau}. \label{sum}
\end{eqnarray}
Below we study the statistics of $W(t)$ as it is obtained by averaging over the statistics of the random strain
$\sigma$ in Eq.~(\ref{evolutionequation}). At large times the statistics of the latter becomes stationary and, by
ergodicity, independent of $\bm x$ (at small times the statistics is generally time-dependent because fixing the initial condition at $t=0$ breaks the translational invariance in time). Since we are interested only in the asymptotic statistics of $\mu_i(t)$ at large times, then, with no loss, we will use for the statistics of $\sigma$ in Eq.~(\ref{evolutionequation}) its large time, stationary, asymptotic form. The stationary statistics holds automatically if we consider
the fluctuations in the steady state when initial points are sampled with the natural measure.
The simplest example of the statistics of $\sigma$ is provided by Kraichnan model where $\sigma$ is Gaussian with zero average and the pair correlation 
\begin{eqnarray}&&
\langle \sigma_{ij}(t_1)\sigma_{mn}(t_2)=\delta(t_2-t_1)\nabla_j\nabla_nK_{im}(0).
\label{Kraichnan}
\end{eqnarray}
Note that when $\sigma$ in Eq.~(\ref{evolutionequation}) is a white noise, still there is no need to specify the interpretation convention (say It\^o or Stratonovich) because the correction term, which arises when changing the interpretation convention, vanishes identically, see \cite{Horvai}.
We now pass to the derivation of the pairing rule.
We follow the same lines as in \cite{Dressler,DettmannMorriss,Ruelle}. We introduce the symplectic tensor $J$ by
\begin{equation}
\label{symplectic}
  J
=
  \begin{pmatrix}
    0 & 1 \\
    -1 & 0 \\
  \end{pmatrix}
,
\end{equation}
that obeys $J^TJ=1$. It is easy to see that $W_s(t)\equiv \exp[-t/\tau]J^T W^{-1, T}(t) J$ obeys
the equation
\begin{eqnarray}&&
\frac{d W_s}{dt}=\begin{pmatrix}
    0 & 1 \\
    \frac{\sigma^T}{\tau} & -\frac{1}{\tau}\\
  \end{pmatrix} W_s, \ \ W_s(0)=I. \label{evolutionequation1}
\end{eqnarray}
Above the subscript $s$ stands for symplectic, see below. Eq.~(\ref{evolutionequation1}) is the same as Eq.~(\ref{evolutionequation}) up to the change $\sigma\to \sigma^T$. Noting that the
eigenvalues of $[W_s^T(t)W_s(t)]^{-1}=\exp[2t/\tau]JW^T(t)W(t)J^T$ coincide with those of
$\exp[2t/\tau]W^T(t)W(t)$, we conclude that the spectrum of $[\ln W_s^T(t)W_s(t)]/2t$ is given by
$-\mu_i(t)-1/\tau$, cf. \cite{Dressler,DettmannMorriss,Ruelle}.

The simplest situation is the one of the potential flow, $\bm u(\bm r, t)=\nabla \phi(\bm r, t)$.
In this case Eqs. (\ref{Newton}) are Hamiltonian up to the linear friction term $-\bm v/\tau$, the
situation considered in \cite{Dressler}. We have $\sigma^T=\sigma$ which leads to $W_s(t)=W(t)$ and
$[\ln W^T(t)W(t)]/2t=[\ln W_s^T(t)W_s(t)]/2t$. The fact that $W$ differs from $J^T W^{-1, T}(t) J$
only by a multiplicative constant is sometimes described by saying that $W$ is conformally symplectic
\cite{Ruelle}. It follows from the above that the spectrum of $[\ln W^T(t)W(t)]/2t$ has a reflection point at $-1/2\tau$ and the so-called strong conjugate pairing rule \cite{Bonnetto} holds: $\mu_i(t)+\mu_{2d+1-i}(t)=-1/\tau$. Such pairing of the spectrum holds for every realization and any time $t$. Taking the limit of large $t$ we find the
pairing of the Lyapunov exponents $\lambda_i+\lambda_{2d+1-i}=-1/\tau$. The relations agree with Eq.~(\ref{sum}).

For turbulence, where vorticity is present, $\sigma^T\neq \sigma$ and we cannot apply the above relations.
Yet, as we pass to show, it is enough for the pairing rule for $\lambda_i$ that $W$ is conformally symplectic
only statistically. If statistics of $\sigma$ is invariant under transposition $\sigma\to \sigma^T$ then
it follows from Eqs.~(\ref{evolutionequation}) and (\ref{evolutionequation1}) that $W(t)$ and
$W_s(t)=\exp[-t/\tau]J^T W^{-1, T}(t) J$ are equal in law. This gives that
\begin{eqnarray}&&
\!\!\!\frac{\ln W^T(t)W(t)}{2t}\simeq \frac{\ln J^T W^{-1}(t)W^{-1, T}(t) J}{2t}-\frac{I}{\tau}, \label{equalityinlaw}
\end{eqnarray}
where we designated the equality in law of random variables by $\simeq$. In the limit $t\to \infty$ the
matrices above have finite limit that produces the pairing rule
\begin{eqnarray}&&
\lambda_i+\lambda_{2d+1-i}=-\frac{1}{\tau}. \label{pairing}
\end{eqnarray}
At finite $t$, Eq.~(\ref{equalityinlaw}) implies a symmetry of the probability density function (PDF)
$P\left(\{\mu_i\}, t\right)$ of $\mu_i(t)$
\begin{eqnarray}&&
P(\mu_1, .. , \mu_{2d})=P\left(-\mu_{2d}-\frac{1}{\tau}, .. , -\mu_1-\frac{1}{\tau}\right). \label{PDFrelation}
\end{eqnarray}
We remind that by our convention on ordering of $\mu_i$ the domain of definition of $P(x_1, .. , x_{2d}, t)$
is $x_1\geq .. \geq x_{2d}$. Let us rewrite Eq.~(\ref{PDFrelation}) in a form more usual for fluctuation relations.
At large times the probability of a finite deviation of $\mu_i$ from their long-time limits $\lambda_i$ decays. It
was shown in \cite{BF} that the decay is exponential when the Lyapunov spectrum is non-degenerate, $\lambda_i>\lambda_{i+1}$, see also \cite{review}. Accounting for the constraint (\ref{sum}) one has
\begin{eqnarray}&&
P\propto
\delta\left(\sum \mu_i+d/\tau\!\right)\exp[-tH(\mu_1, .. , \mu_{2d-1})],
\end{eqnarray}
Here $H(x_1, x_2,.., x_{2d-1})$ is a convex
entropy function which is positive everywhere except for $x_i=\lambda_i$, where it vanishes.
The quadratic expansion of $H$ near its minimum at $x_i=\lambda_i$ would reproduce the central limit theorem
\cite{review}. Using Eq.~(\ref{PDFrelation}) we find the fluctuation relation
\begin{eqnarray}&&
H\left(\mu_1, .. , \mu_{2d-1}\right)=
H\left(\frac{1}{\tau}\!-\!\mu_{2d}, .. ,
\frac{1}{\tau}\!-\!\mu_{2}\right), \label{fluctuationrelation}
\end{eqnarray}
where $\mu_{2d}=-d/\tau-\sum_{i=1}^{2d-1}\mu_i$. Evaluating the above at $\mu_i=\lambda_i$ one recovers Eq.~(\ref{pairing}).

To discuss the relevance of the symmetry $\sigma\to \sigma^T$ for the turbulent flows, let us first note
that since small scale turbulence is isotropic, it is natural to consider the rotationally invariant statistics
of $\sigma$. Then, in Gaussian approximation the statistics of $\sigma$ always has the symmetry $\sigma\to\sigma^T$.
Indeed the mean value and the pair correlation function (fully determining the statistics) have the general form $\langle \sigma_{ij}\rangle=D\delta_{ij}$ and $\langle \sigma_{ij}(t_1)\sigma_{mn}(t_2)\rangle=A(t_2-t_1)\delta_{ij}\delta_{mn}+B(t_2-t_1)\delta_{im}\delta_{jn}
+C(t_2-t_1)\delta_{in}\delta_{mj}$ which is invariant under the substitution $\sigma\to \sigma^T$.
It follows that, in particular, Eq. (\ref{pairing}) holds for Kraichnan model with isotropic statistics.

The fluctuation relation (\ref{PDFrelation}) is different from the Gallavotti-Cohen relation which for PDF
reads as $P(\mu_1, .. , \mu_{2d})=P(-\mu_{2d}, .. , -\mu_1)\exp[\sum\mu_i]$, see \cite{BFF,review}.
Gallavotti-Cohen relation applies under the condition of time-reversibility, which is not obeyed by Eq.~(\ref{Newton}). Nevertheless, under the condition that the "generalized time-reversal" symmetry $\sigma(t)\to \sigma(-t)$ holds, 
one {\it can} derive (\ref{fluctuationrelation}) from Gallavotti-Cohen relation in some transformed
variables. Here "generalized time-reversal" stresses that the dynamics in the phase space is not time-reversible and we talk only about the
symmetry of the strain matrix. We first rewrite Eq.~(\ref{separation}) as
\begin{eqnarray}&&
\frac{d^2\bm R}{dt^2}+\frac{1}{\tau}\frac{d\bm R}{dt}=\frac{\sigma\bm R}{\tau}.
\end{eqnarray}
We introduce $\bm \psi(t)\equiv \exp\left[t/2\tau\right]\bm R$ which satisfies
\begin{eqnarray}&&
\frac{d^2\bm
\psi}{dt^2}=\left[\frac{1}{4\tau^2}+\frac{\sigma}{\tau}\right]\bm
\psi,\ \ \begin{pmatrix}
    \bm \psi(t) \\
    {\dot {\bm \psi}} (t) \\
  \end{pmatrix} 
  ={\tilde
W}(t)\begin{pmatrix}
      \bm \psi(0) \\
    {\dot {\bm \psi}} (0) \\
  \end{pmatrix},\nonumber
\end{eqnarray}
where the evolution matrix ${\tilde W}$ obeys
\begin{eqnarray}&&
\frac{d{\tilde W}}{dt}={\hat m}{\tilde W},\ \ {\hat m}\equiv\begin{pmatrix}
    0 & 1 \\
    \frac{1}{4\tau^2}+\frac{\sigma}{\tau} & 0 \\
  \end{pmatrix} , \ \ {\tilde W}_{ij}(0)=\delta_{ij}.\nonumber
\end{eqnarray}
It is easy to see from the definition of $\bm\psi$ that at large times the finite time Lyapunov exponents
${\tilde \mu}_i(t)$ of ${\tilde W}(t)$ (the eigenvalues of $[\ln {\tilde W}^T(t){\tilde W}(t)]/2t$)
are simply related to $\mu_i(t)$: ${\tilde \mu}_i(t)\approx \mu_i(t)+1/2\tau$.
Then, at large times, Eq.~(\ref{PDFrelation}) is equivalent to the ordinary
Gallavotti-Cohen relation for the PDF of ${\tilde \mu}_i$, that is
$P({\tilde \mu}_1, .. , {\tilde \mu}_{2d})=P(-{\tilde \mu}_{2d}, .. , -{\tilde \mu}_{1})$, where we used
$\det {\tilde W}=1$ and $\sum {\tilde \mu}_i=0$. Thus for the entropy function describing the behavior
at large times, Gallavotti-Cohen relation for ${\tilde W}$ implies the validity of Eq.(\ref{fluctuationrelation}).

Let us prove Gallavotti-Cohen relation for ${\tilde W}$ under the assumption that the statistics of $\sigma(t)$ is invariant under $\sigma(t)\to\sigma(-t)$. 
We introduce the second time argument into $W$ by
\begin{eqnarray}&&
\frac{\partial {\tilde W}(t, s)}{\partial t}={\hat m}(t){\tilde W}(t, s),\ \ {\tilde W}_{ij}(s, s)=\delta_{ij}.
\end{eqnarray}
Writing ${\tilde W}(t, s)$ as the time-ordered exponent
\begin{eqnarray}&&
{\tilde W}\!=\!1\!+\!\int_s^t {\hat m}(t_1)dt_1\!+\!\int_s^t\!dt_1\int_s^{t_1}\!dt_2{\hat m}(t_1){\hat m}(t_2)\!+\!..,\nonumber
\end{eqnarray}
it is easy to see that $\partial_s W(t, s)=-W(t, s){\hat m}(s)$. It follows that
$\partial_sW^{-1}(t, s)={\hat m}(s)W^{-1}(t, s)$. Consequently, $W^{-1}(t)\equiv W^{-1}(t, 0)$
can be found as $M(\tau=t)$ where $\partial_{\tau}M(\tau)=-{\hat m}(t-\tau)M(\tau)$ and $M(\tau=0)=I$,
cf. \cite{review}. Introducing the matrix
\begin{eqnarray}&&
N\equiv  \begin{pmatrix}
    1 & 0 \\
    0 & -1 \\
  \end{pmatrix},
\end{eqnarray}
one finds $\partial_{\tau}N^TM(\tau)N=m(t-\tau)N^TM(\tau)N$. The symmetry $\sigma(t)\to \sigma(-t)$
together with the stationarity imply the symmetry $\sigma(\tau)\to \sigma(t-\tau)$ and 
${\hat m}(\tau)\to {\hat m}(t-\tau)$. We conclude that
the invariance of the statistics of $\sigma$ under the
time-reversal $\sigma(t)\to \sigma(-t)$ implies the equality in law of ${\tilde W}(t)$ and
$N^T {\tilde W}^{-1}(t) N$. This proves the Gallavotti-Cohen relation for ${\tilde W}$, since the
eigenvalues of $N^T {\tilde W}^{-1, T}(t){\tilde W}^{-1}(t) N$ are easily seen to be the same as
of $[{\tilde W}^T(t){\tilde W}(t)]^{-1}$.

Let us repeat that the relation holding as a consequence of $\sigma(t)\to\sigma(-t)$ symmetry is Eq.~(\ref{fluctuationrelation}) and not the Gallavotti-Cohen type symmetry for $\mu_i$: by itself the flow in
the phase space is not time reversible. In fact, it is hard to formulate the conditions on the statistics
of $\bm u(\bm r, t)$ that would ensure the symmetry $\sigma(t)\to \sigma(-t)$. One case where the symmetry holds
is Kraichnan model. Thus the fluctuation relation (\ref{fluctuationrelation})
holds in Kraichnan model always, even if the velocity statistics is not rotationally invariant.

From now on we consider the implications and the applications of the pairing relation for $\lambda_i$. Let us first study the spectrum of the Lyapunov exponents, regarding the latter as a function of particles inertia at fixed
statistics of $\bm u(\bm r, t)$. The inertia can be measured by the dimensionless Stokes number ${\rm St}\equiv \lambda_1^{fl}\tau$, where we introduced the spectrum $\lambda_i^{fl}$, $i=1, 2, .. , d$ of the Lyapunov exponents of the flow $\bm u(\bm r, t)$. At small ${\rm St}$ the effect of inertia is insignificant and the particles simply follow the flow. As a consequence of the pairing rule (\ref{pairing}) we find $\lambda_i\approx \lambda_i^{fl}+O(\tau)$ and $\lambda_{2d+1-i}=-1/\tau-\lambda_i^{fl}+O(\tau)$ for $i=1, 2, .. , d$. The large negative term $-1/\tau$ in $\lambda_i$ with $i>d$ describes fast relaxation of the
velocity to the local velocity of the flow. As ${\rm St}$ increases, eventually one enters the regime where $\lambda_1\tau\gg 1$, which can be called the limit of strong inertia, cf. \cite{HorvaiFouxon}. In this limit one can neglect the $1/\tau$ term in Eq.~(\ref{pairing}) and find the simple reflection symmetry $\lambda_i=-\lambda_{2d+1-i}$. Note that the latter approximation does not respect the identity $\sum \lambda_i=-d/\tau$.

The above can be used to derive a nice expression for Kaplan-Yorke dimension $D_{KY}$ \cite{Ott} of the particles attractor in the phase space. In the limit of large inertia one deals with heavy particles which motion is close to ballistic. As a result, in this limit the dimension of particles' attractor becomes close to the full dimension of the phase space $2d$. Let us see how the dimension gap $C_{KY}=2d-D_{KY}$ closes. Using the reflection symmetry $\sum_{i=1}^{2d-1}\lambda_i\approx \lambda_1>0$ and the identity $\sum \lambda_i=-d/\tau$ one finds that at large
inertia
\begin{eqnarray}&&
C_{KY}\approx \frac{d}{\lambda_1\tau},\ \ \lambda_1\tau\gg 1,
\end{eqnarray}
where we used $\lambda_{2d}\approx -\lambda_1$. In the case of Kraichnan model the scaling argument at large ${\rm St}$ gives $\lambda_1\tau\propto {\rm St}^{1/3}$  \cite{Horvai,Jeremie,HorvaiFouxon} and we find
\begin{eqnarray}&&
C_{KY}\sim {\rm St}^{-1/3},\ \ {\rm St}^{1/3}\gg 1.
\end{eqnarray}
We observe that the decay is rather slow. Moreover, the above analysis implies that $C_{KY}\sim 1$ for
${\rm St}^{1/3}\lesssim 1$. As a result, one expects the particles attractor in the phase space to
have a complex structure up to possibly very large Stokes numbers satisfying ${\rm St}^{1/3}\sim 1$, cf. \cite{HorvaiFouxon}.

The above analysis of the limit of the large inertia has in fact a universal applicability -- Kraichnan model
becomes valid in the limit of large $\tau$ for arbitrary flow with finite correlations in time \cite{Jeremie,HorvaiFouxon}. In this limit the separation of the time-scale $\tau$ of the particle and the
time scales of the flow allows to introduce the effective Langevin description in much the same way as it is done in the ordinary theory of Brownian motion \cite{Ma}. Qualitatively, heavy particles perform effective averaging
of the action of the flow on them, that allows Gaussian approximation, cf. \cite{HorvaiFouxon}.
We make a general conclusion that in the limit of large $\tau$ the codimension of the particles' attractor in
the phase space scales with $\tau$ as $\tau^{-1/3}$. We also conclude that the pairing of the Lyapunov exponents holds always in the limit of large $\tau$.

Another range of situations where Kraichnan model, and thus the pairing, are of universal applicability is the
case where a strong external field is present, such as a constant gravitational or electric field (in the latter
case particles must be charged) \cite{GrishaMisha,HorvaiFouxon,HorvaiFouxon1}. Here the field endows the particles with a finite velocity of the drift through the flow (sedimentation) which becomes large at large field strength. As a result, the particles cross fast the spatial scale of correlations of the velocity gradient and $\sigma(t)$ -- velocity gradient measured in particle's frame -- becomes a fast fluctuating noise describable by Kraichnan model \cite{GrishaMisha,HorvaiFouxon,HorvaiFouxon1}.

Finally, let us discuss the applicability of Kraichnan model to the case where $\bm u(\bm r, t)$ is a developed turbulent flow. The limit of large $\tau$ described above is the limit where $\tau$ is larger than the integral time-scale $t_L$ of turbulence and it is not very practical. Yet, it can be shown
that a modified Kraichnan model holds when $\tau$ is smaller than $t_L$ but larger than the viscous time-scale \cite{HorvaiFouxon,HorvaiFouxon1}. In this version the
amplitude of the noise appearing in Kraichnan model is a random function of time which statistics is not time-reversible. As a result, the relations (\ref{pairing}) and (\ref{fluctuationrelation}) do not hold in general.
However, in the limit ${\rm St}^{1/6}\gg 1$ the model simplifies and one can use as the amplitude not
a random function of time but just a random constant  \cite{HorvaiFouxon,HorvaiFouxon1}. The resulting model already
has the symmetry $\sigma(t)\to \sigma(-t)$. Since for particles in turbulence $\lambda_1\tau\sim {\rm St}^{1/6}$ at
${\rm St}\gg 1$ (but $\tau\ll t_L$) \cite{HorvaiFouxon}, then we conclude that $\lambda_i=-\lambda_{2d+1-i}$ holds
at ${\rm St}^{1/6}\gg 1$. We also conclude that $C_{KY}\sim {\rm St}^{-1/6}$ at ${\rm St}^{1/6}\gg 1$. While the latter
limit is rather theoretical (${\rm St}^{1/6}\gg 1$ and $\tau\ll t_L$ demands very large inertial interval), it allows to make an important conclusion that for particles in turbulence $C_{KY}$ remains of order unity up until very large Stokes numbers, ${\rm St}^{1/6}\sim  1$, see also \cite{HorvaiFouxon}.
Note that here the scaling of $C_{KY}$ with $\tau$ is not $\tau^{-1/3}$ (as claimed above to be universally true in the limit of sufficiently large $\tau$) because of the limitation $\tau\ll t_L$.

Let us summarize. We have shown that two statistical symmetries of the strain matrix, $\sigma\to\sigma^T$ and
$\sigma(t)\to \sigma(-t)$ produce a fluctuation relation for the finite time Lyapunov exponents which differs from Gallavotti-Cohen type relation. In the case of the generalized time-reversal symmetry $\sigma(t)\to \sigma(-t)$, however, the relation follows from Gallavotti-Cohen relation in transformed variables. The fluctuation relation implies that the Lyapunov exponents come in pairs which sum to minus inverse relaxation time of the particles (which is formally the same result as in \cite{Dressler}). We explore the consequences of the pairing of the
Lyapunov exponents in different situations. In particular, we show a universal result that at large inertia the decay of the codimension of the particles' attractor in the phase space is slow -- the codimension decays proportionally
to $\tau^{-1/3}$. For the case of developed turbulent flow with very large inertial interval an intermediate asymptotic regime may hold where the codimension scales as $\tau^{-1/6}$.

We expect that the conjugate pairing rule for the Lyapunov exponents of the inertial particles will be useful
for the numerical measurements of the spectrum of the Lyapunov exponents. The question of the existence of the fluctuation relations for dynamical systems having symmetries other than the time-reversal deserves special 
attention. We also expect that our approach to the derivation of the conjugate pairing rule
for the Lyapunov exponents will find applications to other situations.

We are grateful to K.~Gawedzki for the kind hospitality in Lyon where
the collaboration started. We also thank him for a remark that
helped simplify the exposition of the material.

\end{document}